\documentclass[aps,pre,twocolumn,groupedaddress,showpacs]{revtex4}
\usepackage{graphicx}
\begin{document}

\title{Unfolding a degeneracy point of two unbound states: Crossings
  and anticrossings of energies and widths}
\author{ E. Hern\'andez, A. J\'auregui\dag, \underline{A. Mondrag\'on},
  and L. Nellen\ddag }
\affiliation{Instituto de    F\'{\i}sica, UNAM,
    Apdo. Postal 20-364,  01000 M{\'e}xico D.F., \ M{\'e}xico\\
    $^{\dag}$Departamento de F\'{\i}sica, Universidad de Sonora, 
    Apdo. Postal  1626,  Hermosillo, Sonora, M{\'e}xico \\
$^{\ddag}$Instituto de Ciencias Nucleares, UNAM, Apdo. Postal
    70-543,  04510 M{\'e}xico D.F., \ M{\'e}xico }

\begin{abstract}
We show that when an isolated doublet of unbound states of a physical
system becomes degenerate for some values of the control parameters of
the system, the energy hypersurfaces representing the complex
resonance energy eigenvalues as functions of the control parameters
have an algebraic branch point of rank one in parameter
space. Associated with this singularity in parameter space, the
scattering matrix, $S_{\ell}(E)$, and the Green's function,
$G^{(+)}_{\ell}(k; r,r')$, have one double pole in the unphysical sheet
of the complex energy plane. We characterize the universal unfolding
or deformation of a typical degeneracy point of two unbound states in
parameter space by means of a universal 2-parameter family of
functions which is contact equivalent to the pole position function of
the isolated doublet of resonances at the exceptional point and
includes all small perturbations of the degeneracy condition up to
contact equivalence.
\end{abstract}


\maketitle

\section{Introduction}
Recently, a great deal of attention has been given to the
characterization of the singularities of the surfaces representing the
complex resonance energy eigenvalues at a degeneracy of unbound
states. This problem arises naturally in connection with the
topological phase of unbound states which was predicted by
Hern\'andez, Mondrag\'on and J\'auregui\cite{Hern1,mond1,mond2} and
later and independently by W.D. Heiss\cite{Heiss} and which was
recently verified in a series of beautiful experiments by P. von
Brentano\cite{Brent1,Phil,Brent2} and the Darmstadt
group\cite{Demb1,Demb2}, see also\cite{maily}.

\section{Degeneracy of resonance energy eigenvalues as branch
  points in parameter space\label{sec:titles}}
In this short communication, we will consider the resonance energy
eigenvalues of a radial Schr\"odinger Hamiltonian, $H^{(\ell)}_{r}$,
with a potential $V(r; x_{1},x_{2})$ which is a short ranged function
of the radial distance, r, and depends on at least two external
control parameters $(x_{1},x_{2})$. When the potential $V(r;
x_{1},x_{2})$ has two regions of trapping, the physical system may
have isolated doublets of resonances which may become degenerate for
some special values of the control parameters. For example, a double
square barrier potential has isolated doublets of resonances which may
become degenerate for some special values of the heights and widths of
the barriers \cite{Hern2,van,Hern3}.

In the case under consideration, the regular and physical solutions of
the Hamiltonian are functions of the radial distance, $r$, the wave
number, $k$, and the control parameters $(x_{1},x_{2})$.  When
necessary, we will stress this last functional dependence by adding
the control parameters $(x_{1}, x_{2})$ to the other arguments after a
semicolon.

The energy eigenvalues ${\cal E}_{n}=\bigl(\hbar^{2}/2m\bigr)k^{2}_{n}$ 
of the Hamiltonian $H^{(\ell)}_{r}$ are obtained from the zeroes of the
Jost function, $f(-k; x_{1},x_{2})$ \cite{newton}, where $k_{n}$ is such that
\begin{eqnarray}\label{Jf0}
f(-k_{n};x_{1},x_{2}) = 0.
\end{eqnarray}
When $k_{n}$ lies in the fourth quadrant of the complex $k-$plane,
$Re k_{n} > 0 \hspace{0.3cm}\mbox{and}\hspace{0.3cm}Im k_{n} < 0,$
the corresponding energy eigenvalue, ${\cal E}_{n}$, is a complex
resonance energy eigenvalue.

The condition (\ref{Jf0}) defines, implicitly, the functions
$k_{n}(x_{1}, x_{2})$ as branches of a multivalued function \cite{
  newton} which will be called the wave-number pole position function.
Each branch $k_{n}(x_{1},x_{2})$ of the pole position function is a
continuous, single-valued function of the control parameters. When the
physical system has an isolated doublet of resonances which become
degenerate for some exceptional values of the external parameters,
$(x_{1}^{*},x^{*}_{2})$, the corresponding two branches of the
energy-pole position function, say ${\cal E}_{n}(x_{1},x_{2})$ and
${\cal E}_{n+1}(x_{1},x_{2})$, are equal (cross or coincide) at that
point. As will be shown below, at a degeneracy of
resonances, the energy hypersurfaces representing the complex
resonance energy eigenvalues as functions of the real control
parameters have an algebraic branch point of square root type (rank
one) in parameter space.

{\bf Isolated doublet of resonances:} Let us suppose that there is a
finite bounded and connected region ${\cal M}$ in parameter space and
a finite domain ${\cal D}$ in the fourth quadrant of the complex
$k-$plane, such that, when $(x_{1},x_{2}) \ \epsilon \ {\cal M}$, the
Jost function has two and only two zeroes, $k_{n}$ and $k_{n+1}$, in
the finite domain ${\cal D} \ \epsilon \ {\bf C}$, all other zeroes of
$f(-k;x_{1},x_{2})$ lying outside ${\cal D}$. Then, we say that the
physical system has an isolated doublet of resonances. To make this
situation explicit, the two zeroes of $f(-k;x_{1},x_{2})$,
corresponding to the isolated doublet of resonances are explicitly
factorized as
\begin{equation}\label{jf2}
\begin{array}{c}
f(-k; x_{1},x_{2}) = \Bigl[\bigl(k - \frac{1}{2}(k_{n}+k_{n+1})\bigr)^{2}
- \cr \frac{1}{4}\bigl(k_{n} - 
k_{n+1}\bigr)^{2}\Bigr] 
g_{n,n+1}(k,x_{1},x_{2}).
\end{array}
\end{equation}
When the physical system moves in parameter space from the ordinary
point $(x_{1},x_{2})$ to the exceptional point $(x_{1}^{*},x_{2}^{*})$,
the two simple zeroes, $k_{n}(x_{1},x_{2})$ and
$k_{n+1}(x_{1},x_{2})$, coalesce into one double zero
$k_{d}(x_{1}^{*},x_{2}^{*})$ in the fourth quadrant of the complex
$k-$plane.

If the external parameters take values in a  neighbourhood of
the exceptional point $(x^{*}_{1},x^{*}_{2}) \ \epsilon \ {\cal M}$
and $k \ \epsilon \ {\cal D}$, we may write
\begin{eqnarray}\label{bgr}
g_{n,n+1}(k;x_{1},x_{2})\approx
g_{n,n+1}(k_{d},x^{*}_{1},x^{*}_{2})\neq 0.
\end{eqnarray}
Then,
\begin{eqnarray}\label{cep2}
\Bigl[k -\frac{1}{2}\bigl(k_{n}(x_{1},x_{2}) +
k_{n+1}(x_{1},x_{2})\bigr)\Bigr]^{2} - \cr
\frac{1}{4}\Bigl(k_{n}(x_{1},x_{2}) - k_{n+1}(x_{1},x_{2})\Bigr)^{2}
\approx
\frac{f(-k;x_{1},x_{2})}{g_{n,n+1}(k_{d};x^{*}_{1},x^{*}_{2})},
\end{eqnarray}
the coefficient
$\bigl[g_{n,n+1}(k_{d};x^{*}_{1},x^{*}_{2})\bigr]^{-1}$ multiplying
$f(-k;x_{1},x_{2})$ may be understood as a finite, non-vanishing,
constant scaling factor.

The vanishing of the Jost function defines, implicitly, the pole
position function $k_{n,n+1}(x_{1},x_{2})$ of the isolated doublet of
resonances. Solving eq.(\ref{jf2}) for $k_{n,n+1}$, we get
\begin{eqnarray}\label{knm1}
k_{n,n+1}(x_{1},x_{2}) = \frac{1}{2}\Big(k_{n}(x_{1},x_{2}) +
    k_{n+1}(x_{1},x_{2})\Bigr) + \cr
    \sqrt{\frac{1}{4}\Big(k_{n}(x_{1}-x_{2}\bigr)
      -k_{n+1}\bigl(x_{1},x_{2}\bigr)\Bigr)^{2}}  
\end{eqnarray}
with $(x_{1},x_{2}) \ \epsilon \ {\cal M}$. Since the argument of the
square-root function is complex, it is necessary to specify the
branch. Here and thereafter, the square root of any complex quantity
$F$ will be defined by
\begin{eqnarray}\label{fF}
\sqrt{F} = |\sqrt{F}| \exp\bigl(i\frac{1}{2}arg F\bigr), \hspace{1cm}
0\leq arg F \leq 2\pi
\end{eqnarray}
so that $|\sqrt{F}|= \sqrt{|F|}$ and the $F -$ plane is cut along the
real axis.

Equation (\ref{knm1}) relates the wave number-pole position function
of the doublet of resonances to the wave number-pole position
functions of the individual resonance states in the doublet. 

{\bf The analytical behaviour of the pole-position function at the
  exceptional point}:

The derivatives of the functions $1/2\bigl(k_{n}(x_{1},x_{2}) +
k_{n+1}(x_{1},x_{2})\bigr)$ and $1/4\big(k_{n}(x_{1},x_{2}) -
k_{n+1}(x_{1},x_{2})\big)^{2}$ are finite at the exceptional point.
 They may be computed from the Jost function with the help of the
implicit function theorem \cite{krantz},
\begin{equation}\label{cdknn}
\begin{array}{c}
\Bigl[\Bigl(\frac{\partial\bigl(k_{n}(x_{1},x_{2})-k_{n+1}(x_{1},x_{2})
\bigr)^{2}}{\partial x_{1}}\Bigr)_{x_{2}}\Bigr]_{k=k_{d}}=  \cr
\frac{-8}{\Bigl[\Bigl(\frac{\partial^{2}f(-k;x_{1},x_{2})}{\partial
    k^{2}}\Bigr)_{x^{*}_{1},x^{*}_{2}}\Bigr]_{k=k_{d}} }\Bigl[\Bigl
(\frac{\partial f(-k;x_{1},x_{2})}{\partial
  x_{1}}\Bigr)_{x_{2}}\Bigr]_{k_{d}},
\end{array}
\end{equation}

\begin{equation}\label{dpkx}
\begin{array}{c}
\frac{1}{2}\Bigl[\Bigl(\frac{\partial
  \bigl(k_{n}(x_{1},x_{2})+k_{n+1}(x_{1},x_{2})\bigr)}{\partial
  x_{1}}\Bigr)_{x_{2}}\Bigr]_{k_{d}} = \cr
\frac{-1}{\Bigl[\Bigl(\frac{\partial^{2}f(-k;x_{1},x_{2})}{\partial
    k^{2}}\Bigr)_{x^{*}_{1},x^{*}_{2}}\Bigr]_{k=d_{d}}} 
\Bigl\{\Bigl[\Bigl(\frac{\partial^{2}f(-k;x_{1},x_{2})}{\partial
  x_{1}\partial k}\Bigr)_{x_{2}}\Bigr]_{k=k_{d}} - \cr
\frac{1}{\Bigl[\Bigl(\frac{\partial^{2}f(-k;x_{1},x_{2})}{\partial
    k^{2}}\Bigr)_{x^{*}_{1},x^{*}_{2}}\Bigr]_{k=k_{d}} } 
  \frac{1}{3}\Bigl[\Bigl(\frac{\partial^{3}f(-k;x_{1},x_{2})}{\partial
  k^{3}}\Bigr)_{x^{*}_{1},x^{*}_{2}}\Bigr]_{k=k_{d}} \cr \times \Bigl[\Bigl(
\frac{\partial f(-k;x_{1},x_{2})}{\partial x_{1} }\Bigr)_{x_{2}}
\Bigr]_{k=k_{d}}\Bigr\}.
\end{array}
\end{equation}

From these results, the first terms in a Taylor series expansion of
the functions $1/2\bigl(k_{n}(x_{1},x_{2}) +
k_{n+1}(x_{1},x_{2})\bigr)$ and $1/4\bigl(k_{n}(x_{1},x_{2}) -
k_{n+1}(x_{1},x_{2})\bigr)^{2}$ about the exceptional point
$(x^{*}_{1},x^{*}_{2})$, when substituted in eq.(\ref{knm1}), give
\begin{eqnarray}\label{dpex}
\hat{k}_{n,n+1}(x_{1},x_{2}) = k_{d}(x^{*}_{1},x^{*}_{2}) + \Delta
k_{d}(x_{1},x_{2}) + \cr
\sqrt{\frac{1}{4}\bigl[c^{(1)}_{1}(x_{1}-x^{*}_{1}) 
+  c^{(1)}_{2}(x_{2}-x^{*}_{2})\bigr] }
\end{eqnarray}
for $(x_{1},x_{2})$ in a  neighbourhood of the exceptional point
$(x^{*}_{1},x^{*}_{2})$.  This result may readily be translated into a
similar assertion for the resonance energy-pole position function
${\cal E}_{n,n+1}(x_{1},x_{2})$ and the energy eigenvalues, ${\cal
  E}_{n}(x_{1},x_{2})$ and ${\cal E}_{n+1}(x_{1},x_{2})$, of the
isolated doublet of resonances.

{\bf Energy-pole position function:} Let us take the square of both
sides of eq.(\ref{knm1}), multiplying them by
$\bigl(\hbar^{2}/2m\bigr)$ and recalling ${\cal
  E}_{n}=\bigl(\hbar^{2}/2m\bigr)k^{2}_{n}$, in the approximation of
(\ref{dpex}), we get
\begin{eqnarray}\label{taylor1}
\hat{\cal E}_{n,n+1}(x_{1},x_{2}) &=& {\cal
  E}_{d}(x^{*}_{1},x^{*}_{2}) + 
\Delta {\cal E}_{d}(x_{1},x_{2}) \cr  &+& \hat{\epsilon}_{n,n+1}(x_{1},x_{2}),
\end{eqnarray}
where
\begin{eqnarray}\label{raiz1}
\hat{\epsilon}_{n,n+1}(x_{1},x_{2}) =
\sqrt{\frac{1}{4}\bigl[(\vec{R}\cdot \vec{\xi}) +
 i (\vec{I}\cdot\vec{\xi})\bigr]}\cr
\end{eqnarray}
The components of the real fixed vectors $\vec{R}$ and $\vec{I}$ are
the real and imaginary parts of the coefficients $C^{(1)}_{i}$ of
$(x_{i}-x^{*}_{i})$ in the Taylor expansion of the function
$1/4\bigl({\cal E}_{n}(x_{1},x_{2}) - {\cal
  E}_{n+1}(x_{1},x_{2})\bigr)^{2}$ and the real vector $\vec{\xi}$ is
the position vector of the point $(x_{1},x_{2})$ relative to the
exceptional point $(x^{*}_{1},x^{*}_{2})$ in parameter space.
\begin{eqnarray}\label{xi}
\vec{\xi} = \pmatrix{
\xi_{1} \cr
\xi_{2}
}
= \pmatrix{
x_{1} - x^{*}_{1} \cr
x_{2} - x^{*}_{2}
},
\end{eqnarray}
\begin{eqnarray}\label{RI}
\hspace{0.5cm}\vec{R} = \pmatrix{
Re \ C^{(1)}_{1} \cr
Re \ C^{(1)}_{2}
},
\hspace{0.5cm}\vec{I} = \pmatrix{
Im \ C^{(1)}_{1} \cr
Im \ C^{(1)}_{2}
}.
\end{eqnarray}

The real and imaginary parts of the function
$\hat{\epsilon}_{n,n+1}(x_{1},x_{2})$ are
\begin{equation}\label{rarep}
Re  \hat{\epsilon}_{n,n+1}(x_{1},x_{2}) = \pm
\frac{1}{2\sqrt{2}}\Bigl[{}_{+}\sqrt{\bigl(\vec{R}\cdot\vec{\xi}\bigr)^{2} +
  \bigl(\vec{I}\cdot\vec{\xi}\bigr)^{2}} + \vec{R}\cdot\vec{\xi}\Bigr]^{1/2}
\end{equation}
\begin{equation}\label{raimep}
Im \ \hat{\epsilon}_{n,n+1}(x_{1},x_{2}) = \pm
\frac{1}{2\sqrt{2}}\Bigl[{}_{+}\sqrt{\bigl(\vec{R}\cdot\vec{\xi}\bigr)^{2} +
  \bigl(\vec{I}\cdot\vec{\xi}\bigr)^{2}} - \vec{R}\cdot\vec{\xi}\Bigr]^{1/2}
\end{equation}
and
\begin{eqnarray}\label{signo}
sign \Bigl(Re {\epsilon}_{n,n+1}\Bigr)sign \Bigl(Im
{\epsilon}_{n,n+1}\Bigr) = sign \Bigl(\vec{I}\cdot \vec{\xi}\Bigr)
\end{eqnarray}
It follows from (\ref{rarep}), that $Re 
\hat{\epsilon}_{n,n+1}(x_{1},x_{2})$ is a two branched function of
$(\xi_{1},\xi_{2})$ which may be represented as a two-sheeted surface
$S_{R}$, in a three dimensional Euclidean space with cartesian
coordinates $(Re \hat{\epsilon}_{n,n+1}, \xi_{1},\xi_{2})$. The two
branches of $Re \hat{\epsilon}_{n,n+1}(\xi_{1},\xi_{2})$ are represented
by two sheets which are copies of the plane $(\xi_{1},\xi_{2})$ cut
along a line where the two branches of the function are joined
smoothly. The cut is defined as the locus of the points where the
argument of the square- root function in the right hand side of
(\ref{rarep}) vanishes.  

Therefore, {\it the real part of the energy-pole position function,
  ${\cal E}_{n,n+1}(x_{1},x_{2})$, as a function of the real
  parameters $(x_{1},x_{2})$, has an algebraic branch point of square
  root type (rank one) at the exceptional point with coordinates
  $(x^{*}_{1},x^{*}_{2})$ in parameter space, and a branch cut along a
  line, ${\cal L}_{R}$, that starts at the exceptional point and
  extends in the \underline{positive} direction defined by the unit
  vector $\hat{\xi}_{c}$ satisfying.}
\begin{equation}\label{condi}
\vec{I}\cdot\hat{\xi}_{c} = 0 \hspace{0.5cm}\mbox{and}\hspace{0.5cm}
\vec{R}\cdot\hat{\xi}_{c} = - |\vec{R}\cdot\hat{\xi}_{c}|
\end{equation}

A similar analysis shows that, {\it the imaginary part of the
  energy-pole position function, $Im \ {\cal E}_{n,n+1}(x_{1},x_{2})$,
  as a function of the real parameters $(x_{1},x_{2})$, also has an
  algebraic branch point of square root type (rank one) at the
  exceptional point with coordinates $(x^{*}_{1},x^{*}_{2})$ in
  parameter space, and also has a branch cut along a line, ${\cal
    L}_{I}$, that starts at the exceptional point and extends in the
  \underline{negative} direction defined by the unit vector $\hat{\xi}_{c}$
  satisfying eqs.(\ref{condi})}.

The branch cut lines, ${\cal L}_{R}$ and ${\cal L}_{I}$, are in
orthogonal subspaces of a four dimensional Euclidean space with
coordinates $(Re \epsilon_{n,n+1}, Im  \epsilon_{n,n+1}, \xi_{1},
\xi_{2})$,  but have one point in common, the exceptional point with
coordinates $(x^{*}_{1},x^{*}_{2})$.

The individual resonance energy eigenvalues are conventionally
asociated with the branches of the pole position function according to
\begin{equation}\label{despl1}
\begin{array}{c}
\hat{{\cal E}}_{m}(\xi_{1},\xi_{2}) = {\cal E}_{d}(0,0) + \Delta{\cal
  E}_{n,n+1}(\xi_{1},\xi_{2}) + \cr
\sigma^{(m)}_{R}\frac{1}{2\sqrt{2}}\Bigl[{}_{+}\sqrt{(\vec{R}\cdot
\vec{\xi})^{2} +  (\vec{I}\cdot\vec{\xi})^{2} } +
(\vec{R}\cdot\vec{\xi})\Bigr]^{1/2} + \cr
i\sigma^{(m)}_{I}\frac{1}{2\sqrt{2}}\Bigl[{}_{+}\sqrt{(\vec{R}\cdot
\vec{\xi})^{2} +
  (\vec{I}\cdot\vec{\xi})^{2} } - (\vec{R}\cdot\vec{\xi})\Bigr]^{1/2},
\end{array}
\end{equation}
with $ m = n, n+1$, and
\begin{eqnarray}\label{signat1}
\sigma^{(n)}_{R} = -\sigma^{n+1}_{R}= \frac{Re{\cal E}_{n} - 
Re{\cal E}_{n+1}}{|Re{\cal
    E}_{n} - Re{\cal E}_{n+1}|},
\end{eqnarray}
\begin{eqnarray}\label{signat2}
\sigma^{(n)}_{I} = -\sigma^{n+1}_{I} = \frac{Im{\cal E}_{n} - 
Im{\cal E}_{n+1}}{|Im{\cal
    E}_{n} - Im{\cal E}_{n+1}|}
\end{eqnarray}

Along the line ${\cal L}_{R}$, excluding the exceptional point
$(x^{*}_{1},x^{*}_{2})$,
\begin{eqnarray}\label{rcen}
Re \ {\cal E}_{n}(x_{1},x_{2}) = Re \ {\cal E}_{n+1}(x_{1},x_{2})
\end{eqnarray}
but
\begin{eqnarray}\label{ima}
Im \ {\cal E}_{n}(x_{1},x_{2}) 
\neq Im \ {\cal E}_{n+1}(x_{1},x_{2}).
\end{eqnarray}
Similarly, along the line ${\cal L}_{I}$, excluding the exceptional point,
\begin{eqnarray}\label{icen1}
Im \ {\cal E}_{n}(x_{1},x_{2}) = Im \ {\cal E}_{n+1}(x_{1},x_{2}), 
\end{eqnarray}
but
\begin{eqnarray}\label{rea}
Re \ {\cal E}_{n}(x_{1},x_{2}) 
\neq Re \ {\cal E}_{n+1}(x_{1},x_{2}).
\end{eqnarray}
Equality of the complex resonance energy eigenvalues (degeneracy of
resonances),
${\cal E}_{n}(x_{1}^{*},x_{2}^{*}) = {\cal
  E}_{n+1}(x_{1}^{*},x_{2}^{*}) = {\cal E}_{d}(x^{*}_{1},x^{*}_{2})$,
occurs only at the exceptional point with coordinates
$(x^{*}_{1},x^{*}_{2})$ in parameter space and only at that point.

In consequence, in the complex energy plane, the crossing point of two
simple resonance poles of the scattering matrix is an isolated point
where the scattering matrix has one double resonance pole.

Remark: In the general case, a variation of the vector of parameters
causes a perturbation of the energy eigenvalues. In the particular
case of a double complex resonance energy eigenvalue ${\cal
  E}_{d}(x^{*}_{1},x^{*}_{2})$, associated with a chain of length two
of generalized Jordan-Gamow eigenfunctions \cite{Hern4}, we are
considering here, the perturbation series expansion of the eigenvalues
${\cal E}_{n}, {\cal E}_{n+1}$ about ${\cal E}_{d}$ in terms of the
small parameter $|\xi|$, eqs.(\ref{despl1}-\ref{signat2}), takes the
form of a Puiseux series
\begin{equation}
\begin{array}{c}\label{Tayl2}
{\cal E}_{n,n+1}(x_{1},x_{2}) = {\cal E}_{d}(x^{*}_{1},x^{*}_{2}) +
|\xi|^{1/2}\sqrt{\frac{1}{4}\bigl[(\vec{R}\cdot\hat{\xi}) +
  i(\vec{I}\cdot\hat{\xi})\bigr]} \cr
 + \Delta{\cal E}_{d}(x_{1},x_{2}) + O\Bigl(|\xi|^{3/2}\Bigr)
\end{array}
\end{equation}
with fractional powers $|\xi|^{j/2}, \ j = 0, 1, 2, ...$ of the small
parameter $|\xi|$ \cite{krantz,kato}.

\section{Unfolding of the degeneracy point}
Let us introduce a function $\hat{f}_{doub}(-k;\xi_{1},\xi_{2})$ such that
\begin{eqnarray}\label{hatf}
\hat{f}_{doub}(-k;\xi_{1},\xi_{2}) &=& \Bigl[k -\Bigl(k_{d}(0,0) +
\Delta^{(1)}k_{d}(\xi_{1},\xi_{2})\Bigr)\Bigr]^{2} \cr
&-& \frac{1}{4}\Bigl((\vec{\cal R}\cdot\vec{\xi}) + i(\vec{\cal I}\cdot
\vec{\xi})\Bigr),
\end{eqnarray}
and
\begin{equation}\label{Delta1}
\begin{array}{c}
\Delta^{(1)}k_{d}(x_{1},x_{2}) = \sum_{i =
  1}^{2}d^{(1)}_{i}\xi_{i}
\end{array}
\end{equation}
Close to the exceptional point, the Jost function
$f(-k;\xi_{1},\xi_{2})$ and the family of functions
$\hat{f}_{doub}(-k;\xi_{1},\xi_{2})$ are related by
\begin{equation}\label{jofun}
f(-k; \xi_{1},\xi_{2}) \approx \frac{1}{g_{n,n+1}(k_{d};0,0)}
\hat{f}_{doub}(-k;\xi_{1},\xi_{2})
\end{equation}
the term $\bigl[g_{n,n+1}(k_{d},0,0)\bigr]^{-1}$ may be understood as
a non-vanishing scale factor.

Hence, the two-parameters family of functions
\begin{equation}\label{hjos}
\begin{array}{c}
\hat{f}_{doub}(-k;\xi_{1},\xi_{2}) = \Bigl[k - \Bigl(k_{d}+\Delta^{(1)}k_{d}
(\xi_{1},\xi_{2})\Bigr)\Bigr]^{2} - \cr  \frac{1}{4}\Bigl(\vec{\cal R}
\cdot\vec{\xi} + i\vec{\cal I}\cdot\vec{\xi}\Bigr)
\end{array}
\end{equation}
is contact equivalent to the Jost function $f(-k;\xi_{1},\xi_{2})$ at
the exceptional point. It is also an unfolding \cite{seydel,Poston}  of
$f(-k;\xi_{1},\xi_{2})$ with the following features:
\begin{enumerate}
\item It includes all possible small perturbations of the degeneracy
  conditions
\begin{equation}\label{CO1}
f(-k;\xi_{1},\xi_{2}) = 0, \hspace{0.4cm}
\Bigl(\frac{\partial f(-k;\xi_{1},\xi_{2})}{\partial k}\Bigr)_{k_{d}}
= 0
\end{equation}
\begin{equation}\label{CO3}
\Bigl(\frac{\partial^{2} f(-k;\xi_{1},\xi_{2})}{\partial k^{2}}\Bigr)_{k_{d}}
\neq 0
\end{equation}
up to contact equivalence.
\item It uses the minimum number of parameters, namely two, which is
  the codimension of the degeneracy\cite{mond0}. The
  parameters are $(\xi_{1}, \xi_{2})$.
\end{enumerate}

Therefore, $\hat{f}_{doub}(-k;\xi_{1},\xi_{2})$   {\it is a universal
unfolding \cite{seydel} of the Jost function $f(-k;\xi_{1},\xi_{2})$ at
the exceptional point where the degeneracy of unbound states occurs.

The vanishing of $\hat{f}_{doub}(-k;\xi_{1},\xi_{2})$ defines the
approximate wave number-pole position function }
\begin{equation}\label{gorknn}
\hat{k}_{n,n+1}(\xi_{1},\xi_{2}) = k_{d} +
\Delta^{(1)}_{n,n+1}k_{d}(\xi_{1},\xi_{2}) \pm \Bigl[ 
\frac{1}{4}\bigl(\vec{\cal R}\cdot\vec{\xi} + i\vec{\cal I}
\cdot\vec{\xi}\bigr)\Bigr]^{1/2}
\end{equation}
and the corresponding energy-pole position function $\hat{\cal
  E}_{n,n+1}(\xi_{1},\xi_{2})$ given in eq.(\ref{taylor1}).

Since the functions $\hat{\cal E}_{n}(\xi_{1},\xi_{2})$ and $\hat{\cal
  E}_{n+1}(\xi_{1},\xi_{2})$ are obtained from the vanishing of the
universal unfolding $\hat{f}_{doub}(-k;\xi_{1},\xi_{2})$ of the Jost
function $f(-k;\xi_{1},\xi_{2})$ at the exceptional point, we are
justified in saying that, {\it the family of functions $\hat{\cal
    E}_{n}(\xi_{1},\xi_{2})$ and $\hat{\cal
    E}_{n+1}(\xi_{1},\xi_{2})$, given in eqs.(\ref{despl1}) and
  (\ref{signat1}-\ref{signat2}), is a universal unfolding or
  deformation of a generic degeneracy or crossing point of two unbound state
  energy eigenvalues, which is contact equivalent to the exact
  energy-pole position function of the isolated doublet of resonances
  at the exceptional point, and includes all small perturbations of
  the degeneracy conditions up to contact equivalence }.

\section{Crossings and anticrossings of resonance energies and widths}
Crossings or anticrossings of energies and widths are experimentally
observed when the difference of complex energy eigenvalues ${\cal
  E}_{n}(\xi_{1},\bar{\xi}_{2}) - {\cal
  E}_{n+1}(\xi_{1},\bar{\xi}_{2}) = \Delta E  -i (1/2)\Gamma$ is
measured as function of one slowly varying parameter, $\xi_{1}$,
keeping the other constant, $\xi_{2} = \bar{\xi}^{(i)}_{2}$. A
crossing of energies occurs if the difference of real energies
vanishes, $\Delta E = 0$, for some value $\xi_{1,c}$ of the varying
parameter. An anticrossing of energies means that, for all values of
the varying parameter, $\xi_{1}$, the energies differ, $\Delta E \neq
0$. Crossings and anticrossings of widths are similarly described.

The experimentally determined dependence of the difference of complex
resonance energy eigenvalues on one control parameter, $\xi_{1}$, while
the other is kept constant,
\begin{eqnarray}
\hat{\cal E}_{n}(\xi_{1},\bar{\xi}_{2}^{(i)}) - \hat{\cal
  E}_{n+1}(\xi_{1},\bar{\xi}_{2}^{(i)}) =
\hat{\epsilon}_{n,n+1}(\xi_{1},\bar{\xi}_{2}^{(i)})
\end{eqnarray}
has a simple and straightforward geometrical interpretation, it is
the intersection of the hypersurface
$\hat{\epsilon}_{n,n+1}(\xi_{1},\xi_{2})$
with the hyperplane defined by the condition
$(\xi_{1},\bar{\xi}_{2}^{(i)})$.
\begin{figure}
\begin{center}
\includegraphics[width=230pt,height=240pt]{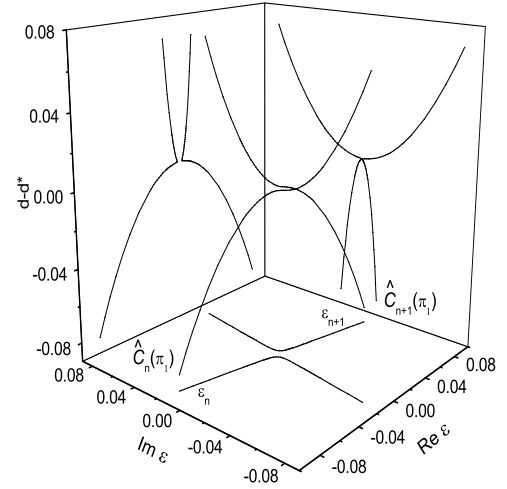}
\caption{The curves $\hat{C}_{n}(\pi_{1})$ and
  $\hat{C}_{n+1}(\pi_{1})$ are the trajectories traced by the points
  $\hat{\cal E}_{n}(\xi_{1},\bar{\xi}_{2}^{(1)})$ and $\hat{\cal
    E}_{n+1}(\xi_{1},\bar{\xi}_{2}^{(1)})$ on the hypersurface
  $\hat{\cal E}_{n,n+1} (\xi_{1},\bar{\xi}_{2}^{(1)})$ when the point
  $(\xi_{1},\bar{\xi}_{2}^{(1)})$ moves along the straight line path
  $\pi_{1}$ in parameter space. In the figure, the path $\pi_{1}$ runs
  parallel to the vertical axis and crosses the line ${\cal L}_{I}$ at
  a point $(\xi_{1,c},\bar{\xi}_{2}^{(1)})$ with $\xi_{1,c} < \xi_{1}^{*}$
  and $\bar{\xi}_{2}^{(1)} < \xi^{*}_{2}$. The projections of
  $\hat{C}_{n}(\pi_{1})$ and $\hat{C}_{n+1}(\pi_{1})$ on the plane
  $(Im {\cal E}, \xi_{1})$ are sections of the surface $S_{I}$; the
  projections of $\hat{C}_{n}(\pi_{1})$ and $\hat{C}_{n+1}(\pi_{1})$
  on the plane $(Re {\cal E}, \xi_{1})$ are sections of the surface
  $S_{R}$.  The projections of $\hat{C}_{n}(\pi_{1})$ and
  $\hat{C}_{n+1}(\pi_{1})$ on the plane $(Re {\cal E}, Im {\cal E})$
  are the trajectories of the $S-$matrix poles in the complex energy
  plane. In the figure, $d-d^{*}=\xi_{1}$ }.
\end{center}
\end{figure}

To relate the geometrical properties of this intersection with the
experimentally determined properties of crossings and anticrossings of
energies and widths, let us consider a point
$(\xi_{1},\bar{\xi}_{2}^{(i)})$ in parameter space away from the
exceptional point. To this point corresponds the pair of
non-degenerate resonance energy eigenvalues ${\cal
  E}_{n}(\xi_{1},\bar{\xi}_{2}^{(i)})$ and ${\cal
  E}_{n+1}(\xi_{1},\bar{\xi}_{2}^{(i)})$, represented by two points on
the hypersurface $\hat{\epsilon}_{n,n+1}(\xi_{1},\xi_{2})$. As the
point $(\xi_{1},\bar{\xi}_{2}^{(i)})$ moves on a straight line path
$\pi_{i}$ in parameter space,
\begin{eqnarray}
\pi_{i} :\hspace{0.2cm} \xi_{1,i} \leq \xi_{1} \leq \xi_{1,f}, \ \ \xi_{2} = 
\bar{\xi}_{2}^{(i)}
\end{eqnarray}
the corresponding points, ${\cal E}_{n}(\xi_{1},\bar{\xi}^{(i)}_{2})$  and
${\cal   E}_{n+1}(\xi_{1},\bar{\xi}_{2}^{(i)})$   trace    two curving
trajectories,  $\hat{C}_{n}(\pi_{1})$ and $\hat{C}_{n+1}(\pi_{1})$  on
the   $\hat{\epsilon}_{n,n+1}(\xi_{1},\xi_{2})$   hypersurface.  Since
$\xi_{2}$ is kept  constant at the fixed value  $\bar{\xi}_{2}^{(i)}$,
the       trajectories     (sections)   $\hat{C}_{n}(\pi_{i})$     and
$\hat{C}_{n+1}(\pi_{i})$,  may be  represented as    three-dimensional
curves  in  a  space ${\cal  E}_{3}$ with   cartesian coordinates $(Re
\epsilon, Im \epsilon, \xi_{1})$, see Figs. 1,2 and 3.
\begin{figure}
\begin{center}
\includegraphics[width=230pt,height=240pt]{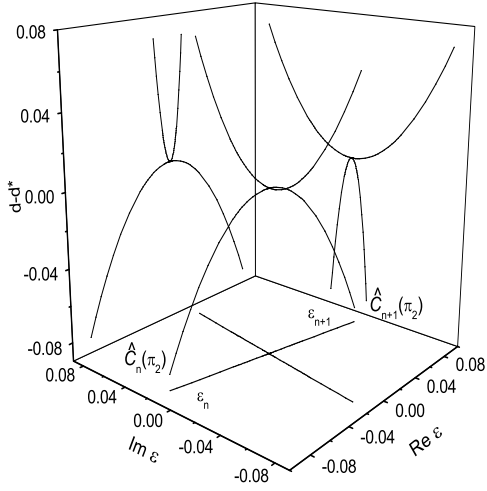}
\caption{The curves $\hat{C}_{n}(\pi_{2})$ and $\hat{C}_{n+1}(\pi_{2})$ are 
  the trajectories of the points $\hat{\cal
    E}_{n}(\xi_{1},\xi^{*}_{2})$ and $\hat{\cal
    E}_{n+1}(\xi_{1},\xi^{*}_{2})$ on the hypersurface $\hat{\cal
    E}_{n,n+1} (\xi_{1},\xi_{2})$ when the point
  $(\xi_{1},\xi^{*}_{2})$ moves along a straight line path $\pi_{2}$
  that goes through the exceptional point $(\xi^{*}_{1},\xi^{*}_{2})$ in
  parameter space. The projections of $\hat{C}_{n}(\pi_{2})$ and
  $\hat{C}_{n+1} (\pi_{2})$ on the planes $(Re {\cal E},\xi_{1})$ and $(Im
  {\cal E},\xi_{1})$ are sections of the surfaces $S_{R}$ and $S_{I}$
  respectively, and show a joint crossing of energies and widths. The
  projections of $\hat{C}_{n}(\pi_{2})$ and $\hat{C}_{n+1}(\pi_{2})$
  on the plane $(Re {\cal E}, Im {\cal E})$ are two straight line
  trajectories of the $S-$matrix poles crossing at 90$^{\circ}$ in the
  complex energy plane. At the crossing point, the two simple poles
  coalesce into one double pole of $S(E)$. }
\end{center}
\end{figure}
\begin{figure}
\begin{center}
  \includegraphics[width=230pt,height=240pt]{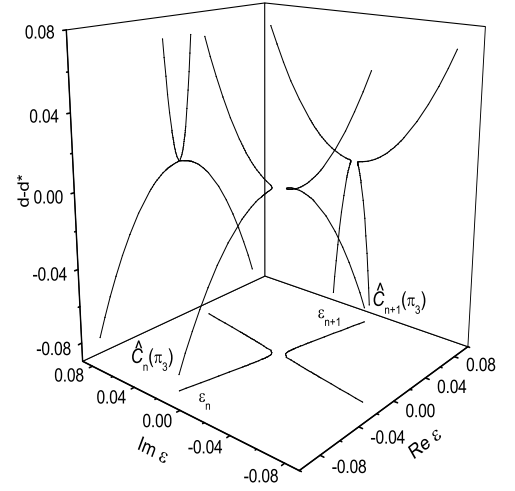}
\caption{The curves $\hat{C}_{n}(\pi_{3})$ and $\hat{C}_{n+1}(\pi_{3})
  $ are the trajectories traced by the points $\hat{\cal E}_{n}
  (\xi_{1},\bar{\xi}^{(3)}_{2})$ and $\hat{\cal
    E}_{n+1}(\xi_{1},\bar{\xi}^{(3)}_{2})$ on the hypersurface ${\cal
    E}_{n,n+1}(\xi_{1},\bar{\xi}_{2}^{(3)})$ when the point
  $(\xi_{1},\bar{\xi}^{(3)}_{2})$ moves along a straight line path
  $\pi_{3}$ going trough the point $({\xi}_{1,c},
  \bar{\xi}^{(3)}_{2})$ with ${\xi}_{1,c} > \xi^{*}_{1}$.  The path
  $\pi_{3}$ crosses the line ${\cal L}_{R}$.  The projections of
  $\hat{C}_{n}(\pi_{3})$ and $\hat{C}_{n+1}(\pi_{3})$ on the plane
  $(Re {\cal E}, \xi_{1})$ show a crossing, but the projections on the
  planes $(Im {\cal E}, \xi_{1})$ and $( Re {\cal E}, Im {\cal E})$ do not
  cross. In the figure, $\xi_{1} = d - d^{*}$.}
\end{center}
\end{figure}
The projections of the curves $\hat{C}_{n}(\pi_{i})$ and
$\hat{C}_{n+1}(\pi_{i})$ on the planes $(Re \epsilon, \xi_{1})$ and
$(Im \epsilon, \xi_{1})$ are  
\begin{eqnarray}\label{Cham}
Re[\hat{C}_{m}(\pi_{i})] = Re \hat{\cal
  E}_{m}(\xi_{1},\bar{\xi}_{2}^{(i)}) \ \ m = n, n+1 
\end{eqnarray}
 and
\begin{eqnarray}\label{Chatm}
Im[\hat{C}_{m}(\pi_{i})] = Im \hat{\cal
  E}_{m}(\xi_{1},\bar{\xi}^{(i)}_{2}) \ \ m = n, n+1
\end{eqnarray}
respectively. 

From eqs.(\ref{despl1}-\ref{signat2}), and keeping
$\xi_{2}=\bar{\xi}^{(i)}_{2}$, we obtain
\begin{equation}\label{delEnEn}
\begin{array}{c}
\Delta E = E_{n}-E_{n+1} = \Bigl(Re\hat{\cal E}_{n} - Re \hat{\cal E}_{n+1}
\Bigr)
\Bigr|_{\xi_{2}=\bar{\xi}_{2}^{(i)}}  \cr
= \frac{\sigma^{(n)}\sqrt{2}}{2}\Bigl[{}_{+}\sqrt{(\vec{R}\cdot\vec{\xi})^{2}+
(\vec{I}\cdot\vec{\xi})^{2}} 
+ (\vec{R}\cdot\vec{\xi})\Bigr]^{1/2}\Bigr]_{\xi_{2}=\bar{\xi}_{2}^{(i)}}
\end{array}
\end{equation}
and 
\begin{equation}\label{delgam}
\begin{array}{c}
\Delta\Gamma = \frac{1}{2}\Bigl(\Gamma_{n} - \Gamma_{n+1}\Bigr) = 
Im\Bigl({\cal E}_{n+1}\Bigr) - 
\Bigl(Im {\cal E}_{n}\Bigr) \cr  = \frac{\sigma^{(n)}_{I}\sqrt{2}}{2}
\Bigl[{}_{+}
\sqrt{(\vec{R}\cdot\vec{\xi})^{2}+(\vec{I}\cdot\vec{\xi})^{2}} - 
(\vec{R}\cdot\vec{\xi})\Bigr]^{1/2}\Bigr|_{\xi_{2}=\bar{\xi}_{2}^{(i)}}
\end{array}
\end{equation}
These expressions allow us to relate the terms $(\vec{R}\cdot\vec{\xi})$
and $(\vec{I}\cdot\vec{\xi})$ directly with observables of the isolated
doublet of resonances. 
Taking the product of  $\Delta E\Delta\Gamma$, and
recalling eq.(\ref{signo}), we get
\begin{eqnarray}\label{delEdelg}
\Delta E\Delta\Gamma = \Bigl(\vec{I}\cdot\vec{\xi}\Bigr)\Bigr|_{\xi_{2}=
\bar{\xi}_{2}^{(i)}}
\end{eqnarray}
and taking the differences of the squares of the left hand sides of
(\ref{delEnEn}) and (\ref{delgam}), we get
\begin{eqnarray}\label{delEsq}
\Bigl(\Delta E\Bigr)^{2} - \frac{1}{4}\Bigl(\Delta\Gamma\Bigr)^{2} = 
\Bigl(\vec{R}\cdot\vec{\xi}\Bigr)\Bigr|_{\xi_{2}=\bar{\xi}_{2}^{(i)}}
\end{eqnarray}

At a crossing of energies $\Delta E$ vanishes, and at a crossing of widths
$\Delta\Gamma$  vanishes.  Hence,  the relation  found in eq.(\ref{delEdelg})
means that {\it a crossing of energies or widths can occur if and only
  if $(\vec{I}\cdot\vec{\xi})_{\bar{\xi}^{(i)}_{2}}$ vanishes}

For a vanishing $(\vec{I}\cdot\vec{\xi}_{c})_{\bar{\xi}^{(i)}_{2}} = 0
= \Delta E\Delta\Gamma$, we find three cases, which are distinguished
by the sign of $(\vec{R}\cdot\vec{\xi}_{c})_{\bar{\xi}^{(i)}_{2}}$.
From eqs. (\ref{delEnEn}) and (\ref{delgam}),
\begin{enumerate}
\item {\it $(\vec{R}\cdot\vec{\xi}_{c})_{\bar{\xi}^{(i)}_{2}} > 0 $ 
implies $\Delta E
      \neq 0$ and $\Delta\Gamma = 0$, i.e. energy anticrossing and width
      crossing}.
  \item {\it $(\vec{R}\cdot\vec{\xi}_{c})_{\bar{\xi}^{(i)}_{2}} = 0 $
      implies $\Delta E = 0$ and $\Delta\Gamma = 0$, that is, joint
      energy and width crossings, which is also degeneracy of the two
      complex resonance energy eigenvalues}.
    \item {\it $(\vec{R}\cdot\vec{\xi}_{c})_{\bar{\xi}^{(i)}_{2}} < 0
        $ implies $\Delta E = 0$ and $\Delta\Gamma\neq 0$, i.e. energy
        crossing and width anticrossing}.
\end{enumerate}
This rich physical scenario of crossings and anticrossings for the
energies and widths of the complex resonance energy eigenvalues,
extends a theorem of von Neumann and Wigner \cite{wigner} for bound
states to the case of unbound states.

The general character of the crossing-anticrossing relations of the
energies and widths of a mixing isolated doublet of resonances,
discussed above, has been experimentally established by P. von
Brentano and his collaborators in a series of beautiful
experiments \cite{Brent1,Phil,Brent2}.

A detailed account of these and other results will be published
elsewhere \cite{hern5,hern6}

\section{Summary and conclusions}
We developed the theory of the unfolding of the energy eigenvalue
surfaces close to a degeneracy point (exceptional point) of two
unbound states of a Hamiltonian depending on control parameters. From
the knowledge of the Jost function, as function of the control
parameters of the system, we derived a 2-parameter family of functions
which is contact equivalent to the exact energy-pole position function
at the exceptional point and includes all small perturbations of the
degeneracy conditions. A simple and explicit, but very accurate,
representation of the eigenenergy surfaces close to the exceptional
point is obtained. In parameter space, the hypersurface representing
the complex resonance energy eigenvalues has an algebraic branch point
of rank one, and branch cuts in its real and imaginary parts extending
in opposite directions in parameter space. The rich phenomenology of
crossings and anticrossings of the energies and widths of the
resonances of an isolated doublet of unbound states of a quantum
system, observed when one control parameter is varied and the other is
kept constant, is fully explained in terms of the local topology of the
eigenenergy hypersurface in the vecinity of the crossing point.

\vspace{0.3cm}
\begin{acknowledgments}
This work was partially supported by CONACyT M\'exico under contract
number 40162-F and by DGAPA-UNAM contract No. PAPIIT:IN116202
\end{acknowledgments}

\end{document}